\documentclass[pra,aps,reprint,notitlepage,superscriptaddress]{revtex4-1}
\usepackage{amssymb}
\usepackage{mathrsfs}
\usepackage{amsmath,amsfonts,bm,graphicx}
\usepackage{color}
\usepackage{etoolbox}
\pdfoutput=1

\newtoggle{arxiv} 
\toggletrue{arxiv}
\iftoggle{arxiv}
{
\pdfoutput=1

}{
\usepackage[colorlinks=true,linkcolor=blue,urlcolor=black,citecolor=blue,dvipdfm]{hyperref}
\usepackage[outdir=./]{epstopdf}

}

\newcommand{\Eq}[1]{Equation (\ref{#1})}
\newcommand{\fig}[1]{Fig. \ref{#1}}
\newcommand{\Fig}[1]{Figure \ref{#1}}

\newcommand{\rvs}[1]{{\textcolor{black}{#1}}}

\newcommand{\vecr}{{\vec{r}}} 
\newcommand{\rc}{{\vec{r}^{~c}}}

\begin{document}

\title{Repetition and pair-interaction of string-like hopping motions in glassy polymers}

\author{Chi-Hang Lam}
\email{C.H.Lam@polyu.edu.hk}
\affiliation{Department of Applied Physics, Hong Kong Polytechnic University, Hung Hom,
Hong Kong, China}
\date{\today }

\begin{abstract}
The dynamics of many glassy systems are known to exhibit string-like hopping motions each consisting of a line of particles displacing one and other. By using molecular dynamics simulations of glassy polymers, we show that these motions become highly repetitive back-and-forth motions as temperature decreases and do not necessarily contribute to net displacements. Particle hops which constitute string-like motions are reversed with a high probability, reaching 73\% and beyond at low temperature. Structural relaxation rate is then dictated not by a simple particle hopping rate but instead by the rate at which particles break away from hopping repetitions. We propose that disruption of string repetitions and hence also structural relaxations are brought about by pair-interactions between strings. 
\end{abstract}

%\pacs{64.70.Q-, 83.80.Sg, 64.70.pj, 61.20.Ja}
% 64.70.Q- 	Theory and modeling of the glass transition
% 83.80.Sg  Polymer melts   
% 64.70.pj  Polymers: glass transitions in  
% 61.20.Ja  Computer modeling and simulation: liquid structure

%\pacs{83.80.Sg, 64.70.pj, 61.20.Ja, 47.15.gm}
% 83.80.Sg  Polymer melts   
% 64.70.pj  Polymers: glass transitions in  
% 61.20.Ja  Computer modeling and simulation: liquid structure
% 47.15.gm  Thin film flows

% 68.35.bm  Polymers: surface structure of
% 36.20.Ey  Molecular dynamics: of macromolecules and polymers
% 68.55     Polymers: flow properties
% 68.55.am  Polymers: film growth
% 68.55.-a, 68.55.J-  Structure: of thin films
% 68.15.+e  Liquid thin films
\maketitle

\section{Introduction}
When cooled sufficiently rapidly, a liquid can be quenched into the glassy state, where motions of the constituent particles become dramatically slowed down resulting in solid-like responses. Despite its prevalence, glass transition remains poorly understood
\cite{biroli2013review,stillinger2013review,berthier2011review}. 
Fundamental questions including the nature of the transition and the mechanism for the dramatic slowing down of the dynamics are still not settled. 
Leading theories of glassy dynamics include Adam-Gibbs theory \cite{adam1965}, random first order transition (RFOT) theory \cite{kirkpatrick1989,wolynes2006}, 
elastic theory \cite{dyre2006review}, mode-coupling theory \cite{gotzebook}, dynamic facilitation \cite{fredrickson1984,palmer1984,ritort2003review,chandler2010review}, and so on.

Molecular dynamics (MD) simulations allow access to the microscopic details of the dynamics \cite{baschnagel2010}. 
A remarkable discovery is string-like motion in which neighboring particles arranged in a line displaces one another from head to tail \cite{glotzer1998}.
They have been observed in simulations of fragile glasses including binary Lennard-Jones liquids \cite{glotzer1998,donati1999spatial}, monoatomic Dzugutov liquids \cite{glotzer2004}, and bead-spring polymers \cite{glotzer2003}. 
They are also observed in experiments of colloidal systems \cite{weeks2000,zhang2011colloid}. 
String-like motions are often considered as structural relaxations, i.e. $\alpha$ relaxations \cite{donati1999spatial}.
\rvs{
They can be classified as coherent or incoherent if the particles involved hop simultaneously or asynchronizely respectively. 
A coherent string-like motion or a coherent part of an incoherent string-like motion is also called a micro-string  \cite{glotzer2004}. 
}
Micro-strings are in general short and most are of 1 to 3 particles long  \cite{glotzer2004}. 
They have been observed to be reversible \cite{glotzer2004,kawasaki2013} and have thus been considered as $\beta$ relaxations \cite{chandler2011}. 
String-like motions extracted from short-time displacements, which are not necessarily coherent, have also been interpreted as $\beta$ relaxations \cite{douglas2013Ni}. 
However, detailed criteria and mechanisms concerning how $\beta$ relaxations are related to structural relaxations is lacking.

An important recent advancement by Chandler and coworkers \cite{chandler2011,isobe2016} is the suggestion of a physical realization of  hierarchically constrained dynamics \cite{palmer1984} based on string-like motions.  The authors identified an elementary excitation as a structural feature allowing a micro-string motion and a larger particle displacement indicates an excitation at a higher level. A simple dependence of the dynamic rate on level was then suggested. The detailed mechanism dictating how each hierarchical level  facilitates or constrains each other deserves further investigation.

Glasses are routinely studied in both equilibrium and non-equilibrium states \cite{wahnstrom1991,sastry1998}. Besides investigating equilibrium samples of a glassy polymer system from MD simulations, we also examine  deeply supercooled non-equilibrium samples quenched to much lower temperature $T$. 
We show that as $T$ decreases, string-like particle hopping motions dominate the dynamics and hopping distances become concentrated around the particle diameter.
We find that string-like motions become increasingly reversible and repetitive. This is directly related to increasingly back-and-forth nature of particle motions commonly observed in glass systems \cite{miyagawa1988,vollmayr2004,vogel2008,ahn2013,helfferich2014}.
We argue that string-like motions, as opposed to only micro-strings, are mostly $\beta$ relaxations at low $T$. Since identical string-like motions can be reversed and repeated multiple times, we refer to them as multiple back-and-forth propagation of a single string. 
We suggest that the main mechanism rendering string-like motions less readily  reversed and may hence contribute to structure relaxations is pair-interactions between strings. 
Pair-interactions also destroy old strings and facilitate the creation of new strings, resulting in effective string mobility.
Examples of string repetitions and interactions are illustrated by particle trajectories from MD simulations. Systematic quantitative assessment of string repetition based on particle hopping statistics are performed.

The rest of this paper is organized as follows. Section \ref{method} describes our simulation method. Section \ref{repetition} illustrates string repetition. Quantitative measurements on particle hoppings and their  repetitive properties are presented respectively in Sec. \ref{distance} and \ref{return}. We explain in Sec. \ref{interaction} the disruption of string repetitions by string interactions. Section \ref{void} discusses the relationship between voids, strings and mobile clusters while Sec. \ref{conclusion} concludes the paper with a summary and some further discussions.

\section{Model and simulation method}
\label{method}

Our simulations are based on a bead-spring model of polymer following Refs. \cite{kremer1990,varnik2002,lam2013crossover} generalized to heavier chain-tails for producing a uniform monomer mobility.
Specifically, pairs of particles interact via the Lennard Jones (LJ) potential 
$4\epsilon \left[\left({\sigma}/{r}\right)^{12} - \left({\sigma}/{r}\right)^{6} \right]$
with a cutoff $2\cdot2^{1/6}$ beyond which it becomes a constant. 
The potential hence has both attractive and repulsive parts. 
Bonded particles are further bounded by a finitely extensible nonlinear elastic (FENE) potential 
$ - \frac{k}{2} R_0^2 \ln \left[ 1 -\left({r}/{R_0}\right)^2 \right]$
where $k=30\epsilon/\sigma^2$ and $R_0=1.5\sigma$. 
We adopt dimensionless LJ units which amounts to taking $\sigma=\epsilon=1$. We nevertheless express lengths in unit of $\sigma$ for clarity.
Internal particles in each chain have a mass $M_I=1$. However, in contrast to common practice \cite{kremer1990,varnik2002,lam2013crossover}, one particle at each end of the chain has a larger mass  $M_T=4$. This leads to a nearly uniform mobility for all particles in the chain as is verified from measured particle mean square displacements (MSD). 

Our main simulations involve two samples of polymer melts each having 1500 chains of length 10 inside a cubic box with periodic boundary conditions simulated using a timestep of 0.005. The melts are first thermalized at $T=0.6$ and then 0.5 following standard techniques \cite{kremer1990}. They are then repeatedly quenched and annealed under NPT conditions at zero pressure by steps of $\Delta T=0.02$ to produce samples at various $T$. Each of these quenching or annealing processes involves $10^7$ timesteps. A sample is further annealed for $10^8$ timesteps before data taking.

We report our main studies at temperature $T$ in between $0.30$ to $0.40$ exhibiting dynamics ranging from those of deeply to slightly supercooled liquids. 
A sample is considered equilibrium if the particle MSD during the final annealing process reaches $5\sigma^2$. For $T=0.36$, 0.38 and 0.40, fully equilibrated polymer samples are studied. For $T=0.30$, 0.32, 0.34 and 0.36, we consider non-equilibrium samples quenched from higher $T$.  Note that at $T=0.36$, we study both equilibrium and quenched samples for comparison. 
Specifically, the sample preparation procedure described above leads to the equilibrium samples for $T=0.38$ and 0.40 and the quenched samples at $T=0.30, ... , 0.36$. For $T=0.36$, we further perform an extended annealing of $10^{10}$ timesteps and this generates two additional samples at $T=0.36$ satisfying the above equilibrium criterion.

All our analysis are based on coarse-grained particle trajectories $\rc_i(t)$ 
recorded during data collection runs each consisting of $10^9$ additional simulation timesteps  under NVT conditions. Specifically, each value of $\rc_i(t)$ is a coarse-grained position of particle $i$ defined by
\begin{equation}
  \rc_i(t) = \langle ~ \vec{r}_i(t') ~ \rangle _ {t' \in [t, t+\Delta t_c] }
\end{equation}
averaged over a duration $\Delta t_c= 5$ (i.e. 1000 timesteps). In practice, this is evaluated by an in-situ average over 20 instantaneous positions $\vec{r}_i(t')$ of particle $i$ taken after every 50 timesteps. 
The coarsening time $\Delta t_c= 5$  is long enough to average out most vibrations about meta-stable positions. It is also negligible compared to the typical waiting time between two consecutive hops of a particle. Therefore, $\rc_i(t)$ nearly always points to a meta-stable particle position, rather than somewhere interpolating between two meta-stable positions related by a hop. This improves the accuracy of our measured hopping statistics.
For each sample, we record $10^4$ snapshots of coarse-grained positions $\rc_i(t)$ taken after every duration of $\delta t=500$ (i.e. $10^5$ timesteps). Note that $\delta t$ is still much shorter than the average waiting time between particle hops so that motions neglected in between snapshots only have small quantitative impacts on our results. 

Quantitative measurements to be presented are averaged over two independent samples, for which error bars are generally smaller than the symbol sizes in our plots. They are based on $\rc_i(t)$ and are in general consistent with those using $\vec{r}_i(t)$ except at length scales much smaller than $\sigma$ where the reduced fluctuations may lead to quantitative impacts.
Simulations are executed using the highly efficient HOOMD package \cite{hoomd} on Nvidia GTX580 GPU's. Each extended annealing at $T=0.36$ takes 5 months of runtime. 
Our MD simulations at $T=0.36$ each involving more than $10^{10}$ timesteps is in our knowledge far more intensive than similar previous studies.
Equilibration at $T\le 0.34$ will be  substantially more challenging computationally and should be impractical with current computing technology. 

\section{String repetition}
\label{repetition}

\begin{figure}[tb]
\includegraphics[height=1.2in]{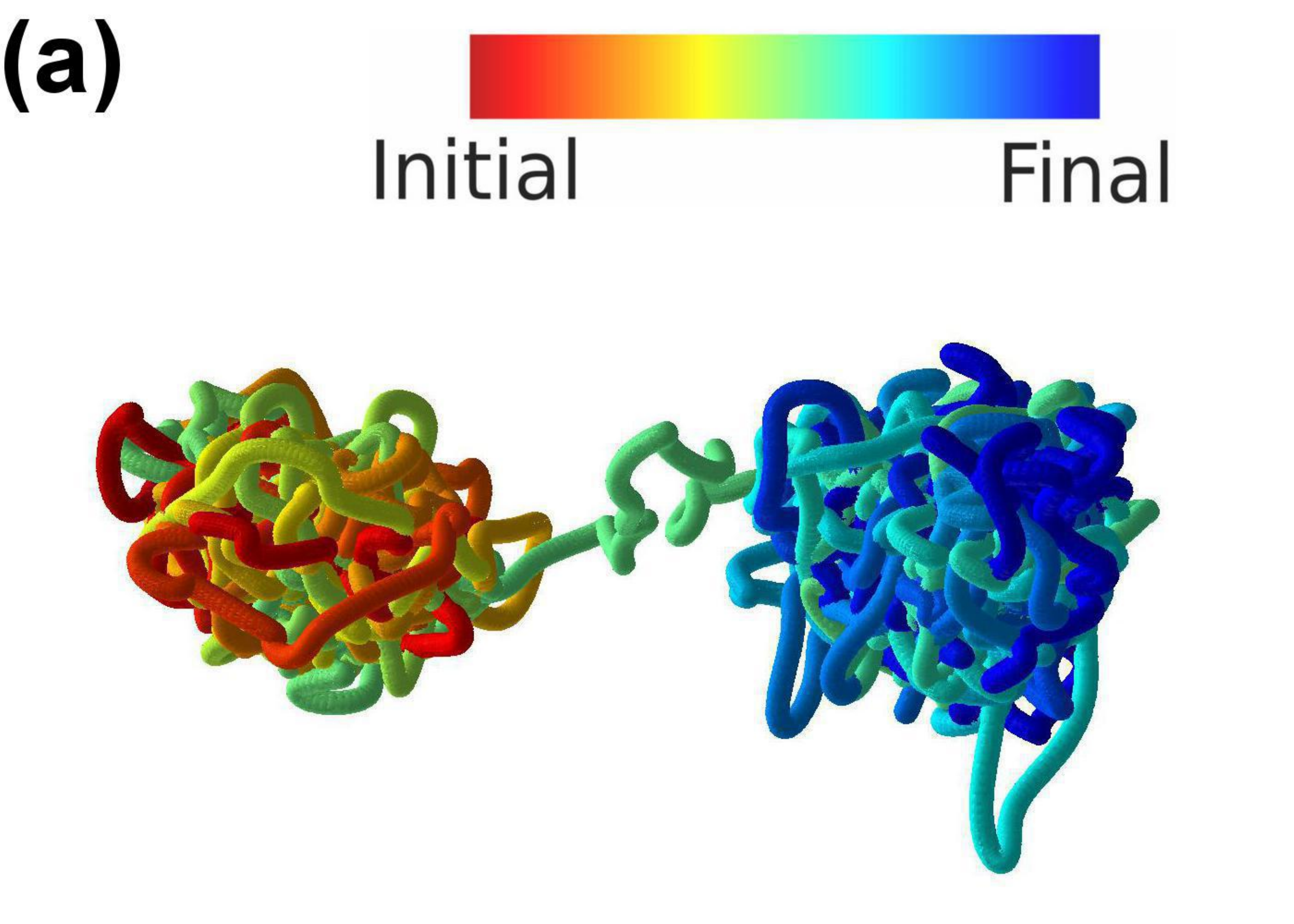} ~~~
\includegraphics[height=1.2in]{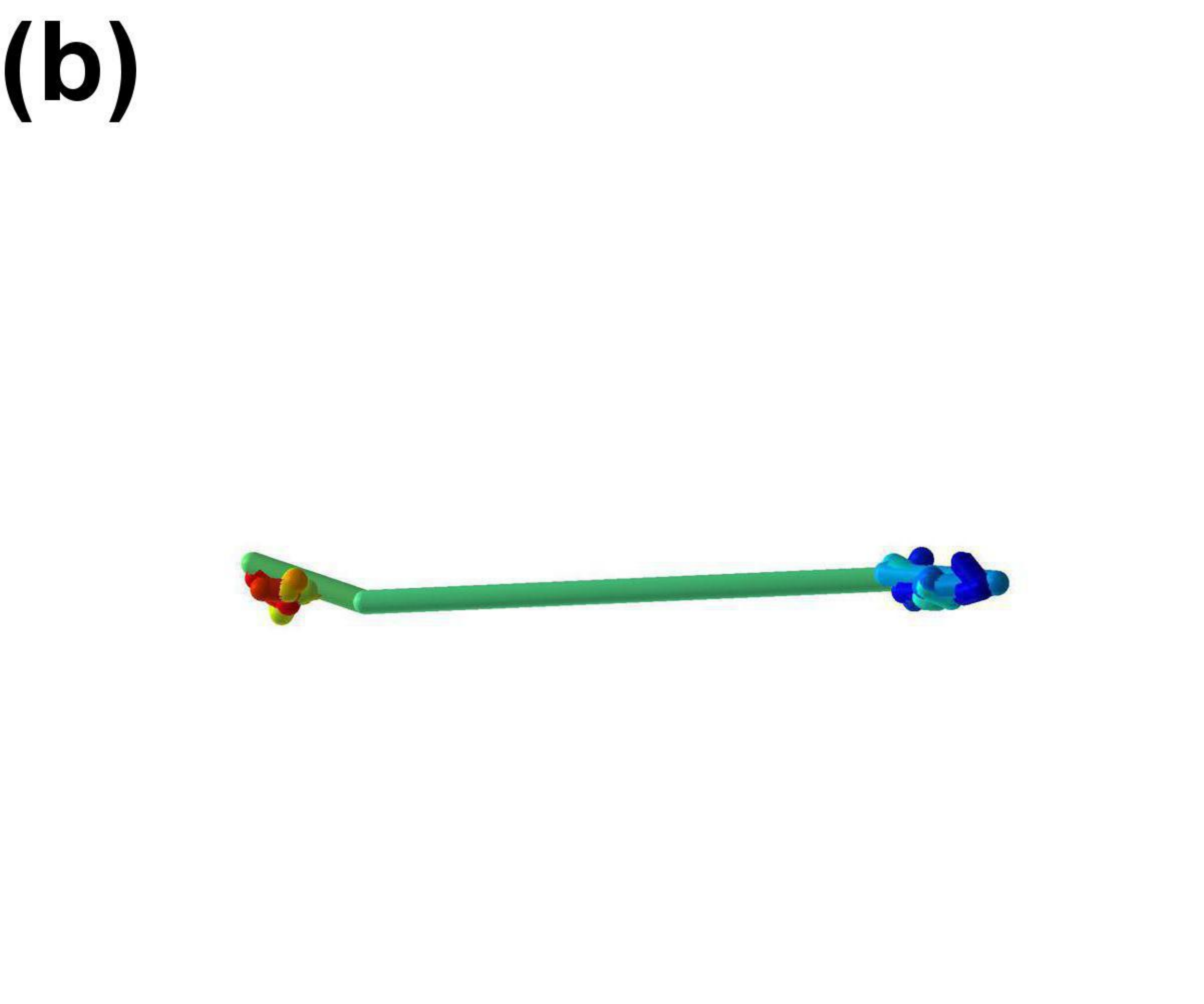} \\
\includegraphics[width=3.1in]{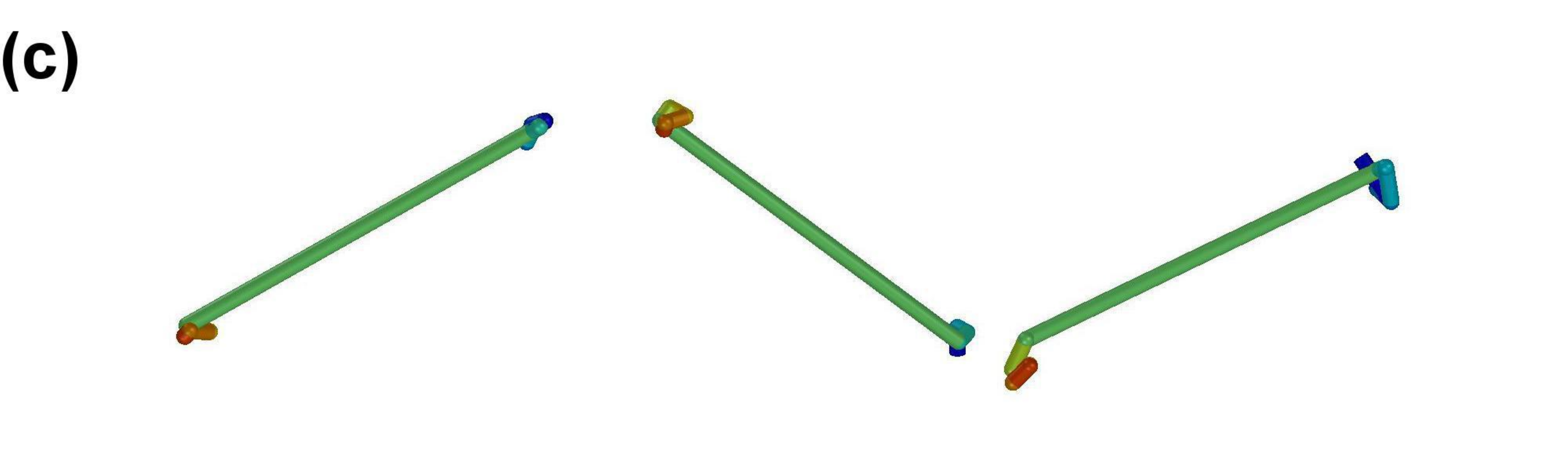} ~~~~~\\
\includegraphics[width=3.1in]{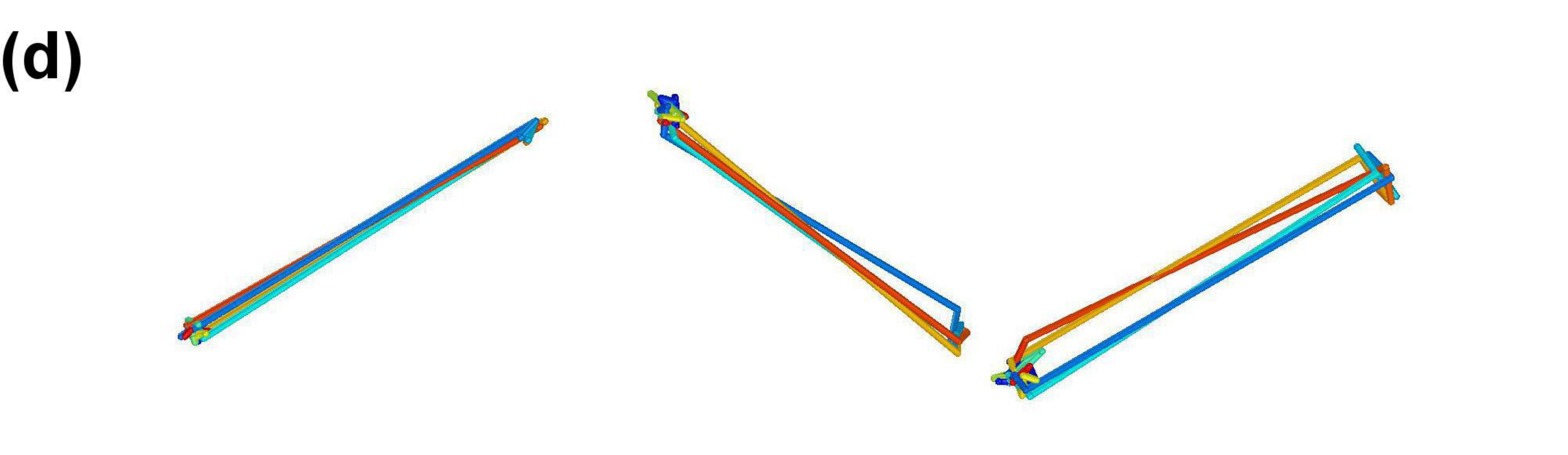} ~~~~~\\
\caption{Particle trajectories colored according to time $t$ with red (blue) being the earliest (latest). (a) Hopping of a single particle as shown by its true trajectory $\vec r_i(t)$. The particle first vibrates about one site and then hops to another site to the right. 
(b) The same motion as in (a) represented by the coarse-grained trajectory consisting of line segments joining time-averaged positions $\rc_i(t)$. Vibrations before (red) and after (blue) the hop are deemphasized by the averaging. The long line segment (green) of length about $0.9\sigma$ implies a sudden large displacement, i.e. a hop. 
(c) Coarse-grained trajectories of hopped particles with displacements larger than $0.6\sigma$ in a neighborhood, showing 3 particles performing a string-like motion. Particles hop to the right and nearly fill the positions of the  preceding particles. The uniform color (green) of the 3 long segments implies simultaneous hops. (d) Coarse-grained trajectories of the same particles in (c) shown for a longer duration. Four string-like motions (red, orange, light blue, blue) occur, the first of which is the one in (c).
The durations shown are 200 in (a)-(b), 5,000 in (c) and 20,000 in (d), while $T=0.32$ in LJ units.
}
\label{Figstr}
\end{figure}

At low temperature $T$, particle motions mainly consist of thermal vibrations and occasional hops.
Figure \ref{Figstr}(a)-(b) shows a typical hop of particle $i$ from a quenched sample at $T=0.32$ as represented by the true trajectory $\vec{r}_i(t)$ recorded at every MD timestep and the coarse-grained trajectory $\rc_i(t)$ explained in Sec. \ref{method}.
As seen, the coarse-grained trajectory clearly illustrates the hopping motion and will be adopted throughout this work.
We consider particle $i$ hopped during a given period if the recorded $\rc_i(t)$ shows any  displacement beyond a distance of $0.6\sigma$ within $any$ sub-interval of the period. In particular, a hopped particle may have performed a round trip and attain a zero net displacement.

It is instructive to visually identify string-like motions directly from the trajectories. Figure \ref{Figstr}(c) plot the trajectories of all hopped particles in a neighborhood. We observe three particles performing a typical string-like motion. Note that the particles involved in general do not belong to the same polymer chain \cite{glotzer2003}. Neighboring particles that have not hopped are not shown.

A key observation in this work is that string-like motions become highly repetitive exhibiting strong memory effects at low $T$. For the string-like motion shown in Fig. \ref{Figstr}(c), it is subsequently reversed, repeated and reversed again as shown in Fig. \ref{Figstr}(d). There are thus 4 successive string-like motions.

\begin{figure}[bt]
\includegraphics[width=0.44\textwidth]{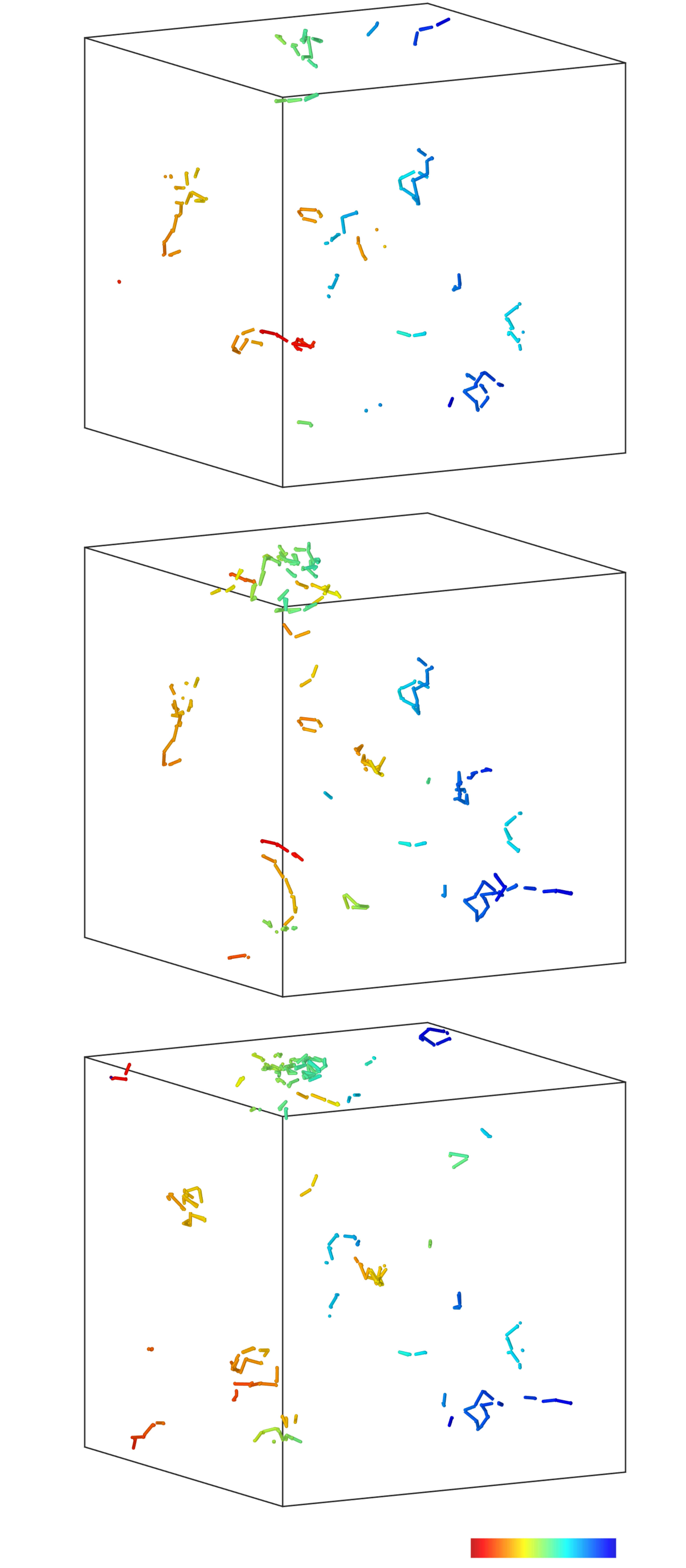}
\caption{ Coarse-grained trajectories of hopped particles during three consecutive durations each of 2500 from an equilibrium sample at $T=0.36$.  Displacements are colored from red to blue starting from the left surface of the cubic simulation box of width $24.3\sigma$. }
\label{Figwhole36}
\end{figure}

\begin{figure}[bt]
\includegraphics[width=0.44\textwidth]{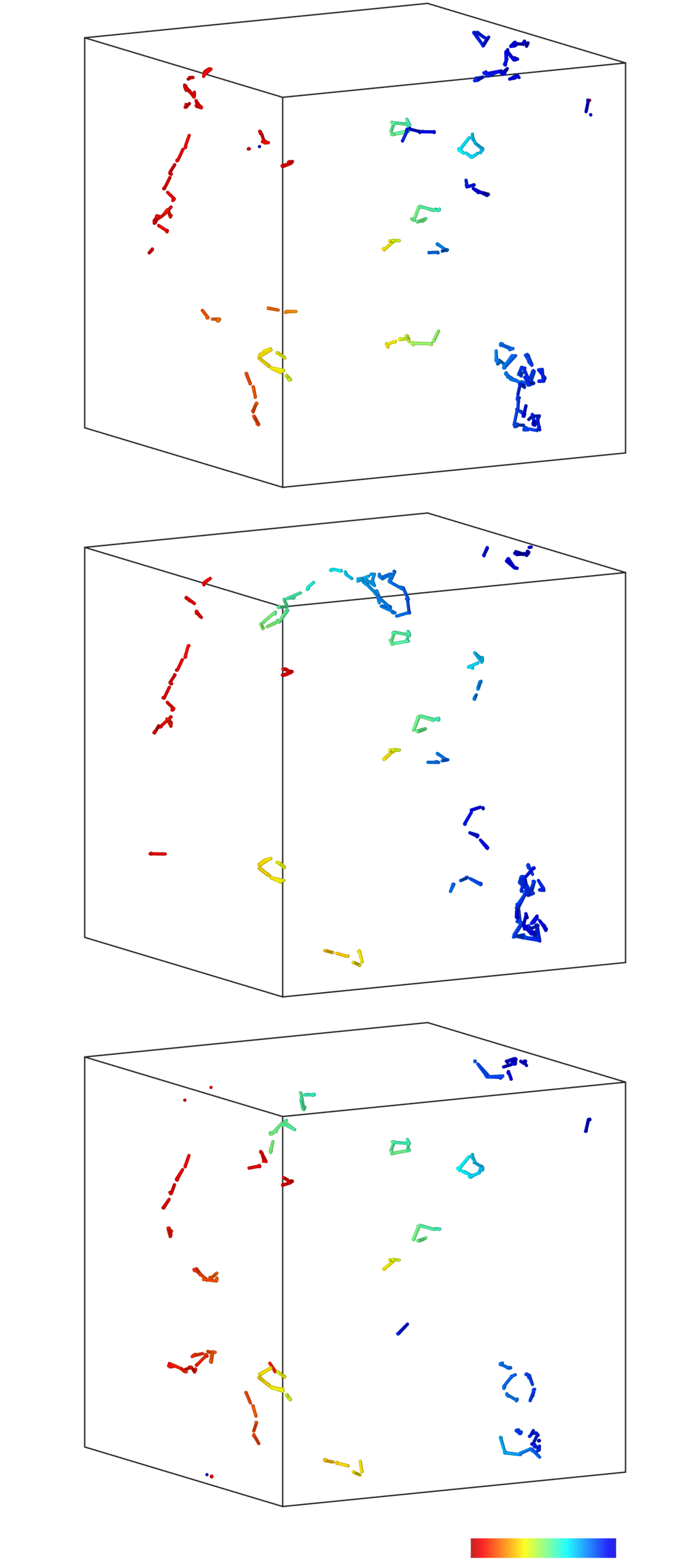}
\caption{
Coarse-grained trajectories of hopped particles during three consecutive durations each of 5000 from a quenched sample at $T=0.32$. The coloring is similar to that in Fig. \ref{Figwhole36}. }
\label{Figwhole32}
\end{figure}

Figures \ref{Figwhole36} and \ref{Figwhole32} show time-consecutive coarse-grained trajectories of all hopped particles from an equilibrium sample at $T=0.36$ and a quenched sample at $T=0.32$ respectively. Line segments constituting the trajectories are colored according to their positions in lateral direction. This allows an indication of depth, as seemingly spatially overlapping trajectories of distinct colors must be well separated in three dimensions. We can observe dozens of examples of string repetition. We have also produced movies showing the evolution of the trajectories for much longer durations \cite{movie}.

From various visualizations, we observe repetitions of both coherent and incoherent string-like motions of different lengths. Incoherent strings can be either fully or partially reversed and repeated. We also provide an animation of the repetition of an incoherent string-like motion \cite{movie}. Persistent repetitions over long durations usually occur to those far away from the other strings. Repetitions become increasingly common and robust as $T$ decreases.

String-like motions at higher $T$ are often considered structural relaxations \cite{donati1999spatial}. 
Observing here that they may not contribute to net displacements, these oscillatory parts of the motions should be classified only as caging dynamics or $\beta$ relaxations alongside thermal vibrations. We will explain later that the dominant mechanism to break away from the indefinite repetitions of string-like motions is a pair interaction between strings.  After repetitions finally stop and assume an odd number of hops, only the last outstanding hop of each particle contributes to structural relaxations. 

String-like motions can be easily identified visually from particle trajectories obtained from MD simulations as shown above or from potential energy landscape (PEL) calculations \cite{swayamjyoti2014}. 
Conventionally, they are often extracted using standard methods which also dictate their precise working definitions \cite{glotzer2003}.
 Algorithmically, the net displacements of particles over a time period of $\Delta t$ comparable to an average structural relaxation time are typically studied. Particles with the top 5\%  displacements  during $\Delta t$ are grouped into clusters of adjacent mobile particles. Each cluster is then resolved into string-like motions on the basis that particles belonging to a string best displace the preceding members. 

We have also identified and quantitatively analyzed string-like motions following conventional algorithms \cite{glotzer2003} adapted to consider particles with displacements beyond $0.6\sigma$ over relatively short time periods $\Delta t$ to allow applications to low $T$. String lengths are found to nicely follow an exponential distribution down to the string length of 1 particle with an average string length $l \simeq 2$ particles varying only mildly with $\Delta t$ and $T$. This is consistent with recent results on Ni nano-particles \cite{douglas2013Ni}. By definition, all hops belong to some string-like motions as a string can be one particle long. In the following, we will simply focus on particle hopping statistics which have straightforward implications on string statistics.

\section{Hopping distance}
\label{distance}
\newcommand{\Probr}{4\pi r^2 G_s(r,\tau)}

\begin{figure}[tb]
\includegraphics[width=0.48\textwidth]{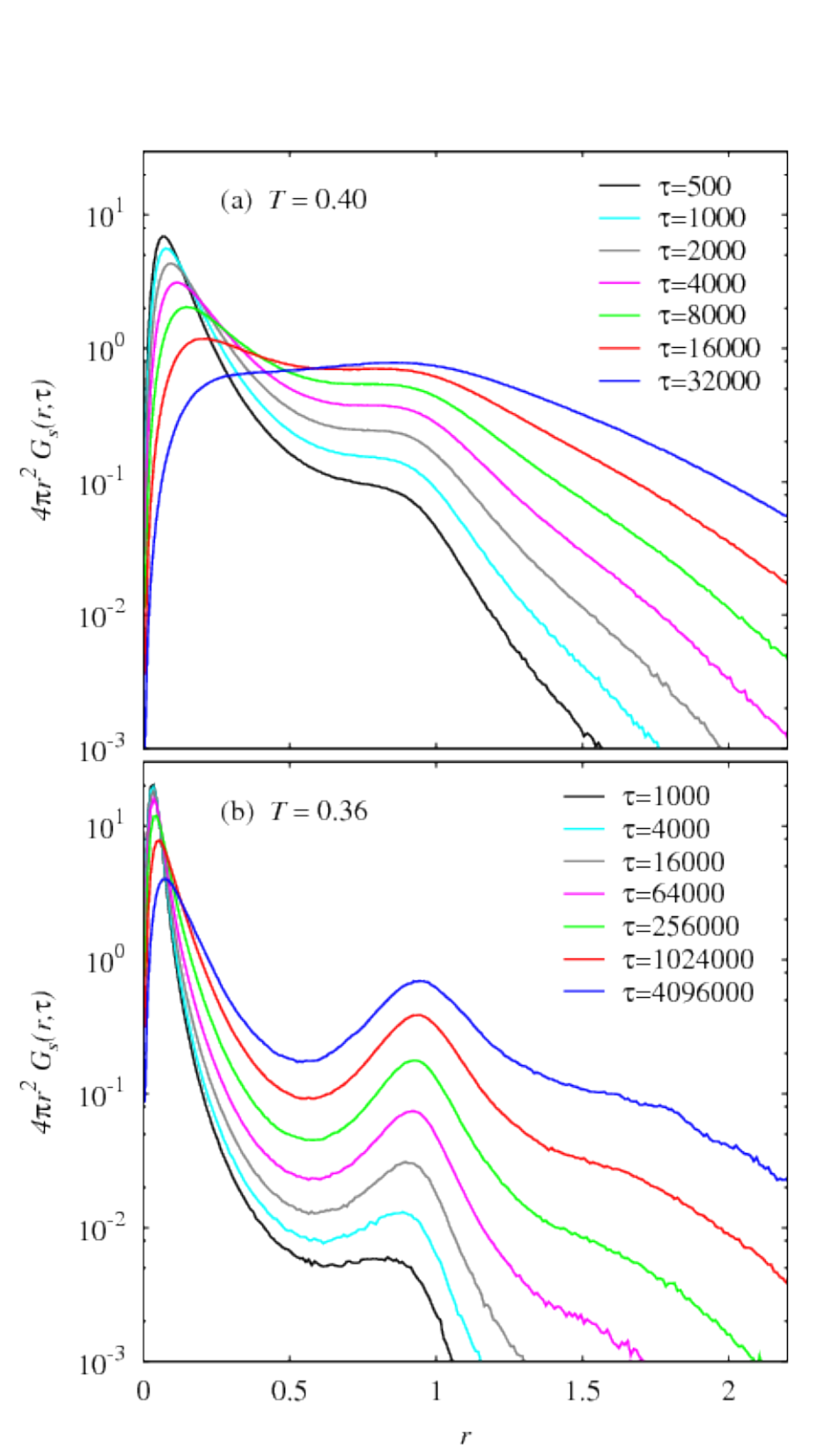} ~~~
\caption{Semi-log plot of probability density $P_\tau(r)=\Probr$ of particle displacement $r$ during time $\tau$ at temperature $T=0.40$ (a) and 0.36 (b) measured from equilibrium samples.}
\label{histtau}
\end{figure}

\begin{figure}[tb]
\includegraphics[width=0.48\textwidth]{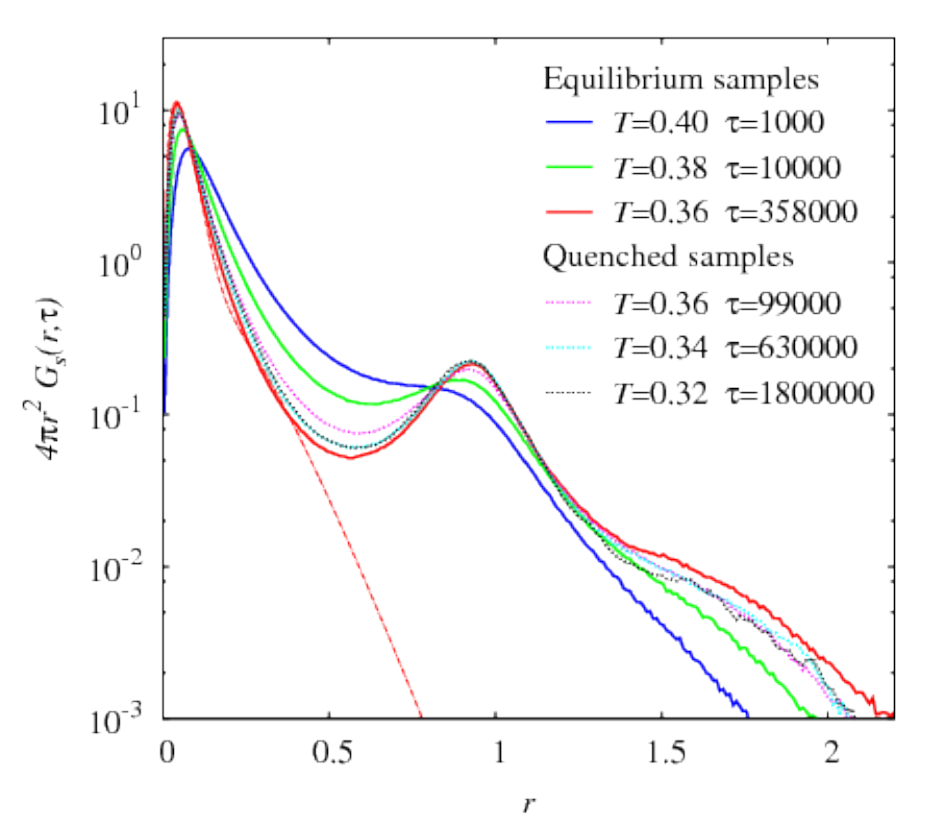} ~~~
\caption{Semi-log plot of probability density $P_\tau(r)=\Probr$ of particle displacement $r$ during time $\tau$ corresponding to a MSD of $0.09\sigma^2$. The dashed line indicates a fit to the main peak for the equilibrium case at $T=0.36$. }
\label{hist}
\end{figure}

We define particle displacement $r$ over time $\tau$ based on coarse-grained trajectories as
\begin{equation}
 r = ~ \mid\rc_i(t_0+\tau)- \rc_i(t_0) \mid .
\end{equation}
We calculate the probability density $P_\tau(r)$ of $r$, which is often written  as $P_\tau(r) = \Probr$ using the van Hove self-correlation function \cite{wahnstrom1991} % $G_s(r,\tau)$ defined as
\begin{equation}
  G_s(r,\tau) = \frac{1}{N} \sum_{i=1}^N \langle \delta(\rc_i(t_0+\tau)- \rc_i(t_0) -\vecr ) \rangle
\end{equation}
where $N$ is the total number of particles.
\Fig{histtau}(a) plots $\Probr$ for $T=0.40$. 
At small $\tau$, it shows a main peak at $r \simeq 0.1\sigma$. It corresponds to  small displacements resulting from both structural relaxations as well as vibrations that are not completely eliminated by the time-coarsening process. As $\tau$ increases, the peak broadens gradually as expected from  relaxations induced by liquid-like collective flow. The distribution however deviates significantly from a Gaussian with a weaker secondary peak at $r \simeq 0.9 \sigma$ resulting from activated particle hops like those shown in \fig{Figstr}. 
Similar secondary peaks have long been observed for binary Lennard Jones liquids  \cite{wahnstrom1991,sastry1998,kawasaki2013} and glassy colloidal liquids \cite{marcus1999} 

\Fig{histtau}(b) shows $\Probr$ at $T=0.36$ for the fully equilibrated samples. The secondary peak at $0.9\sigma$ is now much more prominent. As $\tau$ increases,  it gains in height dramatically. In contrast, the main peak broadens relatively mildly over the same range of $\tau$. The importance of hopping as compared to collective flow is thus significantly increased at this lower $T$. Note that hopping also induces broadening of the main peak as neighboring non-hopping particles are displaced due to longer-ranged elastic distortions. Therefore, collective flow may have diminished even more than naively suggested by the slow broadening of the main peak.

\Fig{hist} compares  $\Probr$  at various $T$ with $\tau$ chosen so that the MSD in all cases are standardized at 0.09$\sigma^2$, a value attainable for our data at $T\ge 0.32$. From the equilibrium samples, we observe a monotonic strengthening of the secondary peak as $T$ decreases implying enhanced importance of hopping as compared to collective flow. 
Comparing between equilibrium and quenched cases at $T=0.36$, we find that equilibration enhances the relative importance of hopping. The quenched samples at lower $T$ however show a weaker $T$ dependence and this may be due to the structural properties of the samples being frozen at $T$ around 0.36. 
In additions, there is a broader peak at $r \simeq 1.6\sigma$ corresponding to double hops of particles. Similar peaks have been reported before for binary Lennard Jones liquids \cite{sastry1998}.

\renewcommand{\xi}{r}
We further study the equilibrium case at $T=0.36$ in detail. The main peak can be fitted to the empirical form 
$f(\xi)=200 \xi^2 \left( 260 e^{-50 \xi} + 200 e^{-15 \xi} \right )$ shown by the dashed line in \Fig{hist}. This portion of the distribution contributes 
$\int \xi^2  f d\xi \simeq 0.0103 \sigma^2$ to the MSD 
and accounts for collective flow, elastic distortion due to particle hops, and vibrations not completely coarsened out. 
The complementary distribution $4\pi r^2G_s-f$ capturing the secondary peak in contrast contributes $0.0797 \sigma^2$ to the MSD and accounts for particle hops. As a result, particle hops are responsible for at least 87\% of the MSD and clearly dominate the dynamics. This dominance in the MSD is even stronger at larger $\tau$ as can be more readily observed from plots of $4\pi r^4 G_s(r,\tau)$ \cite{kawasaki2013}.
We therefore believe that particle hops of distances centered around $\sigma$ completely dominate structural relaxations at low $T$.

From \fig{histtau} and \fig{hist}, the dominant hopping distance of about $0.9\sigma$ is quite independent of $\tau$ and $T$. 
Being close to $\sigma$, it is also close to the distance between meta-stable particle positions and is thus favored. 
A further subtle point is that being slightly smaller than $\sigma$, a particle usually slightly fall short of reaching a neighboring metastable site as revealed by a gap between trajectories of adjacent particles in string-like motions (see \fig{Figstr}(c)). This may indicate some elastic restoring forces counteracting the hop. Finally, $0.6\sigma$ marks the approximate position of the dip which is also quite independent of $\tau$ and $T$. It is adopted as the threshold displacement for defining hops as used above and throughout this work. In contrast, we find that the conventional approach of inferring a threshold displacement from the top 5\% most mobile particles \cite{glotzer2003} may not always best distinguish between the two peaks observed here.

\section{Hopping rates and return probability}
\label{return}

\begin{figure}[tb]
 \includegraphics[width=0.48\textwidth]{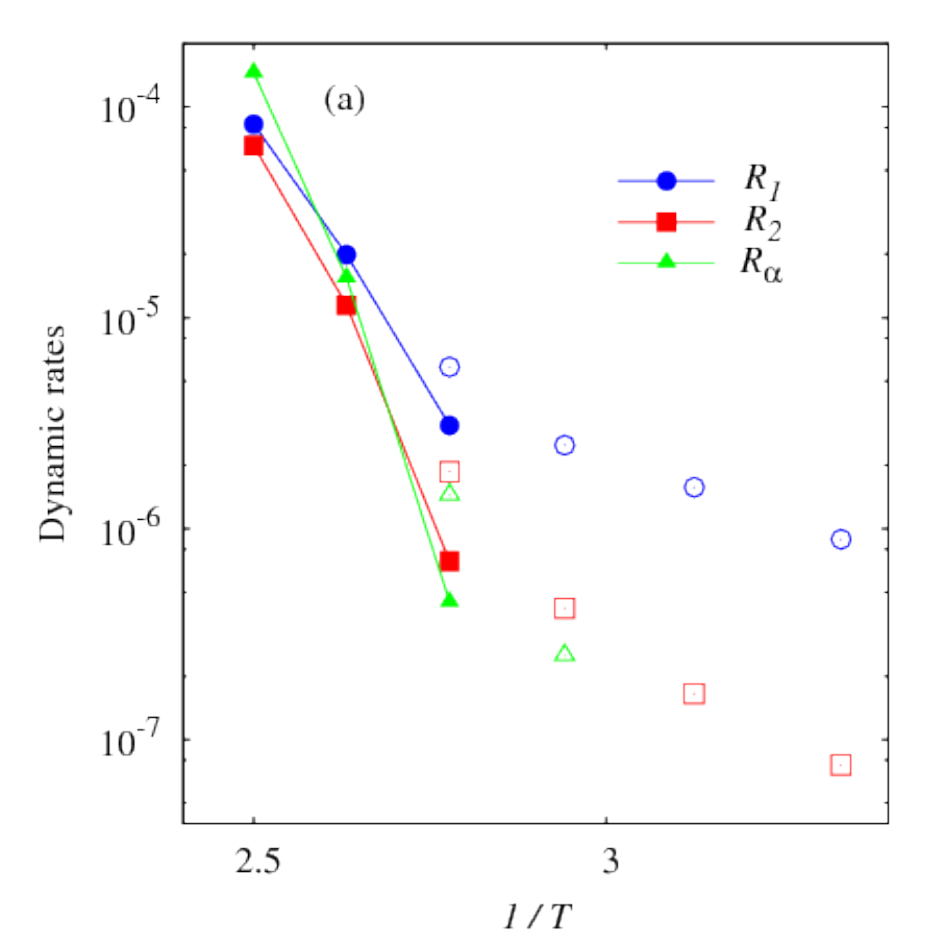}
 \includegraphics[width=0.48\textwidth]{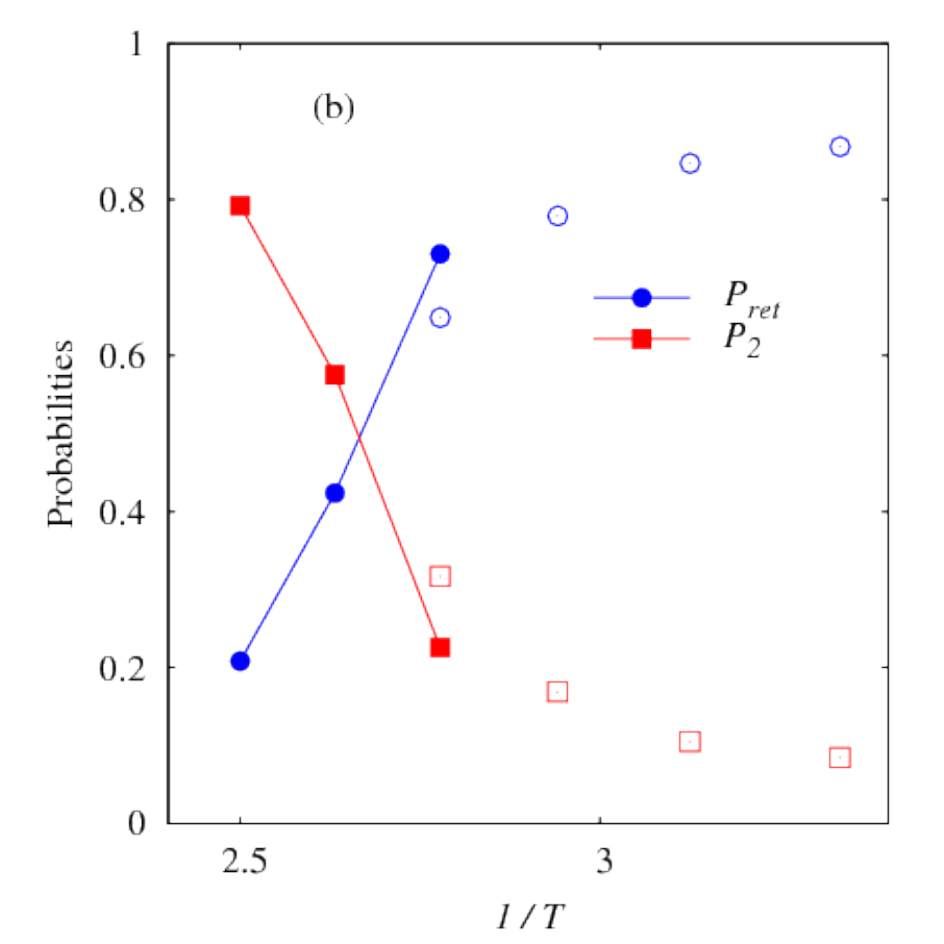}
\caption{(a) Plot of particle hopping rate $R_1$, rate of non-returning second hop $R_2$, and MSD-based structural relaxation rate $R_\alpha$ vs $1/T$ for $T=0.40, 0.38, ... , 0.30$. (b) Plot of  probabilities of returning hop $P_{ret}$ and non-returning second hop $P_2$. In (a) and (b), results based on equilibrium (quenched) samples are represented by solid (open) symbols.
}
\label{Figrate}
\end{figure}

We calculate the particle hopping rate $R_1= P_1/\delta t$ using the probability $P_1$ that particle $i$ hops during $[t_0, t_0+\delta t]$ where $\delta t=500$. Figure \ref{Figrate}(a) plots $R_1$ against $1/T$. The equilibrium data is consistent with a slightly super-Arrhenius behavior. The non-equilibrium data show more moderate decrease of $R_1$ as $T$ is decreased further.

We now focus on the subsequent motion of a hopped particle $i$ by evaluating the probabilities $P_{ret}$ and $P_{2}$ that it will first perform, respectively, a returning or non-returning second hop. Specifically, after the first hop during $[t_0, t_0+\delta t]$, a second hop is defined as returning if the particle returns to within a distance of $0.3\sigma$ from the original position $\rc_i(t_0)$. A non-returning second hop means that it has displaced again elsewhere beyond a distance of $0.6\sigma$ from the hopped position $\rc_i(t_0+\delta t)$. 
We monitor the particle up to a long duration of $10^6$ after the first hop so that subsequent motion can be categorized in most cases. 
Figure \ref{Figrate}(b) plots $P_{ret}$ and $P_{2}$ against $1/T$. As $T$ decreases, $P_{ret}$ rises while $P_2$ drops implying increasing memory effects. We find that $P_{ret}+P_{2} \agt 0.95$ so that uncategorized cases are few. 
These non-returning second hops are responsible for the broader peak of $\Probr$ at  $r \simeq 1.6\sigma$ as observable from \fig{hist}.

More importantly, $P_{ret}$ reaches a high value of 0.73 from our equilibrium data at $T=0.36$. It is well beyond the value of 0.21 found at $T=0.40$ more akin to  simple diffusion. For the quenched samples at $T=0.36$, $P_{ret}$ takes a slightly smaller value of 0.65, which is expected as the quenched samples may partially exhibit properties of higher-temperature structures not completely annealed away. For $T\le 0.34$, the non-equilibrium values of $P_{ret}$ shown in Fig \ref{Figrate}(b) are hence likely the lower bounds of the equilibrium ones which cannot be directly measured. 
We find that  $P_{ret}$ continues to increase as $T$ decreases and reaches a very high probability of 0.93 when we extend our study to $T=0.26$ (data not shown).
This leads us to postulate that for both equilibrium and quenched samples,
\begin{equation}
  \label{pret}
  P_{ret} \rightarrow 1 \mbox{~~~~ and ~~~~} P_2 \rightarrow 0
\end{equation}
as $T \rightarrow 0$.

Figure \ref{Figrate}(a) also plots the rate $R_2$ of non-returning second hops defined by
\begin{equation}
  \label{R2}
  R_2 = R_1 P_2.
\end{equation}
The equilibrium data shows that $R_2$ is slightly super-Arrhenius, like $R_1$. However, due to the hopping repetitions, $R_2$ clearly drops faster than $R_1$ as $T$ decreases. This trend is also demonstrated by the non-equilibrium data at lower $T$. 

In addition, we have measured a MSD-based structural relaxation 
time $\tau_{\alpha}$  \cite{baschnagel2007} which is the time it takes for the particle MSD to reach $(0.6\sigma)^2$.
This leads to a structural relaxation rate defined by
\begin{equation}
 R_{\alpha}=1/\tau_{\alpha} .
\end{equation}
We plot $R_{\alpha}$ in Fig. \ref{Figrate}(a) for $T\ge 0.34$ at which the MSD is able to reach $(0.6\sigma)^2$ during the data collection runs.
For $T \ge 0.38 $, Fig. \ref{Figrate}(a) reveals $R_1 \simeq R_2 \simeq R_{\alpha}$ implying that most hops constitute structural relaxations. In sharp contrast, for $T \le 0.36$ for both equilibrium and quenched cases, $R_1 \gg R_{\alpha}$. It  shows that particle hops,  which have become increasingly repetitive, 
are more frequent than the rate of structural relaxations. This supports the assertion that oscillatory hops are $\beta$ relaxations. On the other hand, $R_2 \simeq R_{\alpha}$ still holds. Therefore, the second non-returning hops, which break string repetitions, now dictate the structural relaxation rate. 

\section{String interaction}
\label{interaction}

\begin{figure}[tb]
 \includegraphics[width=1.5in]{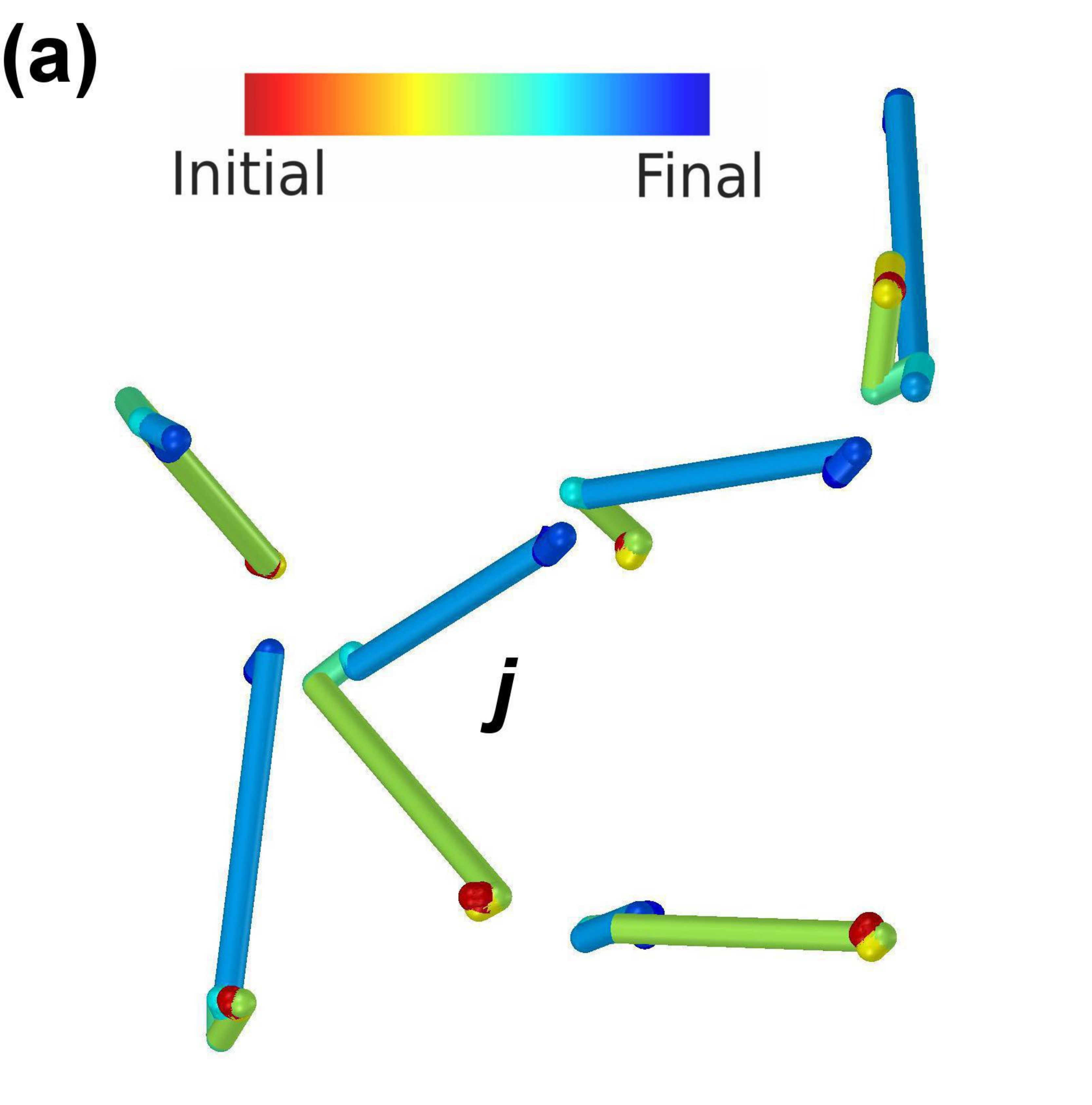}
 \includegraphics[width=1.5in]{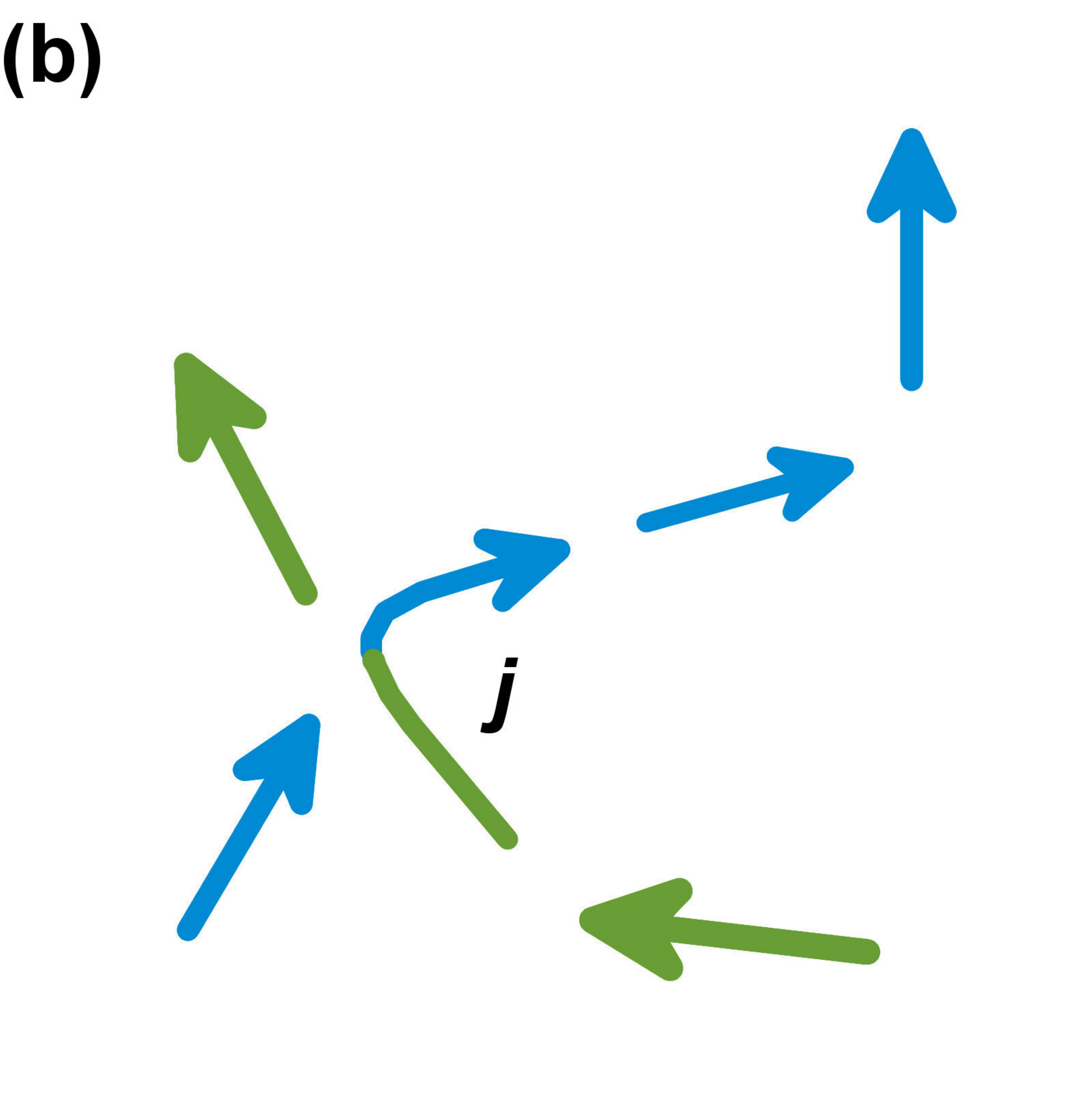}
\caption{
Displacement pathways of 6 particles during a pair-interaction between two strings illustrated using (a) coarse-grained trajectories from a MD simulation and (b) the corresponding schematic diagram. Trajectories in (a) covering a duration of 4500 are colored according to time similar to those in Fig. \ref{Figstr}. The two distinct colors (green and blue) of the 7 long segments imply two set of hops performed at different times. The first string (green) involves the hopping of 3 particles. It pushes particle $j$ to a new position along the second string (blue) whereby the 4 particles the second string contains can hop subsequently. Notice that particle $j$, which is shared by both strings, has hopped twice while each of the other 5 particles have hopped only once. The temperature is $0.32$.
}
\label{Figint}
\end{figure}

To understand the mechanism responsible for the non-returning second hops, we have visually examined hundreds of relevant particle trajectories. Interpretation for their origin get significantly easier at low $T$ as string-like hopping motions completely dominate over collective flow.  For $T \alt 0.32$, all non-returning second hops can be identified as being caused by pair-interactions of strings similar to the example of string crossing shown in Fig. \ref{Figint}. Here, a particle labeled as $j$ is shared by two strings. The first string-like motion (green) induces the first hop of particle $j$. The new position of $j$ enables a new string (blue). The second string-like motion along the new string then induces the second hop of particle $j$.  This however disable the first string so that it cannot be reversed unless the second string is first reversed.  If not reversed eventually, all 7 hops contribute to structural relaxation.

We thus conclude from our observations that individual string repetitions at low $T$ occur indefinitely contributing only to $\beta$ relaxations. Pair interactions can break the repetition and generate structural relaxations. At slightly higher $T$, this mechanism can still be observed but are sometimes complicated by the presence of liquid-like collective flow.

Note that conventional algorithms \cite{glotzer2003} outlined in Sec. \ref{repetition} fail to properly resolve the strings in Fig. \ref{Figint}.  Considering only the net displacement, conventional algorithms overlook the turn made by particle $j$. With the turn inappropriately straightened up, particle $j$ is instead assigned according to the best-displacing rule to an upward-going string of 4 particles on the right. The other 2 particles constitute another upward going string on the left. The algorithms thus report two non-intersecting incoherent strings. This  
hides the detailed time sequence of the dynamics and the phenomenon of string interaction.

We have found it straightforward to identify any specific string in ways consistent with the dynamics by visually inspecting time-colored particle trajectories taken over some manually selected time-windows. However, a practical algorithm remains to be worked out.   

\section{Void, String, Cluster and Cluster repetition}
\label{void}

Our understanding of the relationship between conventionally defined strings and mobile clusters \cite{glotzer2003} based on close examinations of particle trajectories from MD simulations is as follows. For a specific period $\Delta t$, a number of string-like motions and pair interactions may have occurred. Using conventional algorithms that  examine only the net displacements of hopped particles, strings propagated once or repeated an odd number of times lead to the observations of the associated string-like motions. On the other hand, particles in strings repeated an even number of times are classified as immobile since they return to the original positions. The associated string-like motions are not observed.
Furthermore, pair interactions of strings can lead to cross-links between string-like motions. Those observed string-like motions in general cross-linked constitute the mobile clusters.

As explained above, string-like motions become increasingly repetitive as $T$ decreases and repetitions can be broken by string interactions. In fact, as $T$ further decreases, we observe that pair interactions can also be reversed and repeated. More generally, we observe reversal and repetition of sequences of string-like motions and pair-interactions, and thus also of mobile clusters.  An example of a partially repeated cluster of strings shaded in blue can be seen at the lower right corner of the simulation box in Fig. \ref{Figwhole32}.
Here, we note that the repetitive nature of strings and more generally clusters are due to the momentarily frozen self-generated disorders characteristic of  glassy fluids. The recurring clusters exemplify memory effects at longer time scales. 
Structural relaxations then require breaking the repetition of clusters by larger sets of string interactions.

Free volumes are long known to be important in glassy dynamics \cite{cohen1961}.  Due to  hard-core intermolecular potentials, local spikes in the particle density are unlikely and free volumes characterize most density fluctuations.  A contiguous free volume comparable to the average particle volume $\Omega$ can be called a void. A particle hop of a distance about $\sigma$ is equivalent to the hop of a void over the same distance in the reversed direction.  However, it is easy to see from simple geometry that a shorter hop of a particle does not mean a shorter hop of a void. Instead, 
a particle hop of displacement say $0.5\sigma$ corresponds to the hop of half a void over a distance $\sigma$. Since hops are dominated by distances of about $\sigma$ as shown in Sec. \ref{distance}, voids are transported mostly in whole. Voids thus provide a useful decription of the dynamics we have observed.

Nevertheless, static particle configurations show that contiguous voids of volume $\Omega$ are rare.
Smaller voids each of volume a fraction of $\Omega$ are observed in simulations but are correlated only slightly to hopping energy barriers \cite{swayamjyoti2014}.  Voronoi volumes of individual particles have been used to indicate free volumes but they also show only small correlations with motions \cite{starr2002,weitz2005,harrowell2006}. 

To understand the apparent conflict about the relevance of voids as suggested by the dynamics versus the statics,
we have examined particle trajectories around string-like motions in detail.
Since we define particle hops with a threshold displacement $0.6\sigma$, strings like that in  \fig{Figstr}(c) terminates abruptly. More detailed examinations indicate that on each end of a string, there are typically a few  neighboring particles with smaller displacements decreasing continuously further away from each end of the string. The free-volume $\Omega$ transported by a hop is hence distributed across a few interstitial sites before and after the hop. From static observations, their existence may be easily overshadowed by  structural disorder. Detection using Voronoi particle volumes  can be even more challenging as each interstitial volume is further shared among the  surrounding particles. In addition, a free volume may induce a hop and then be transported away, further weakening the spatial correlation between particle hops and free-volumes.  

Therefore, we define a free volume $\Omega$ transported by a hop and distributed  among a few neighboring interstitial sites as a fragmented void. It is a spatially non-contiguous quasi-particle of a combined volume $\Omega$ dominated by a small void a fraction of $\Omega$ at one end of a micro-string accompanied by a few even smaller voids nearby. A fragmented void admits a free energy cost and is thus a structural excitation. We propose that they are the elementary excitations enabling micro-string propagations, which are the elementary dynamical process \cite{chandler2011}. 
Also, a fragmented void is required for a micro-string propagation and a micro-string transports its fragmented void from one of its ends to the other. 
A few micro-strings can share a fragmented void producing incoherent strings of linear or branched geometries.
Since a fragmented void is dominated by a smaller void with a size a fraction of $\Omega$, in the following, we will refer to a fragmented void simply as a void.

We find that particle hopping dynamics can be understood most easily via the motions of voids. 
As a non-trivial example, we observe that in an incoherent string consisting of three micro-strings at $T=0.32$, the micro-strings propagate and reverse repeatedly in a correlated manner most easily described by a single void transversing randomly along the incoherent string. An animation showing the motion is provided \cite{movie}.
As a second example, the mobile cluster of the two micro-strings in \fig{Figint} requires two voids to properly understand.
As collective flow subsides at low $T$, such descriptions become increasingly applicable. At  $T \alt 0.32$, the only examples of micro-strings apparently unrelated to voids are loops, which are comparatively few in number and will be neglected for simplicity. 

We intuitively understand glassy dynamics at low $T$ as dominated by string-like motions and repetitions as follows. A void is typically trapped by the rugged energy landscape to within $n+1$ adjacent meta-stable sites with similar energies that are lower than those of the surrounding sites. The thermally activated discrete random walks of the void among these lower-energy sites are equivalently repetitive string-like motions of the $n$ particles. 
More precisely, as micro-strings can consist of two or more particles, the random walks of a void in general comprise of groups of short-range correlated hops along the string. After a void hops, the local energy landscape experienced by another void nearby may be perturbed. This realizes a string interaction.
The motions of a collection of voids mutually interacting by altering the energy landscape of the others generate the mobile clusters and thus the glassy dynamics. 

\section{Conclusions}
\label{conclusion}

In summary, string-like motions in a typical glassy fluid are studied by MD simulations. As $T$ is decreased, particle hops of distances comparable to the particle diameter are shown to dominate the dynamics. 
These hops and the string-like motions they constitute become increasingly repetitive at low $T$. We propose the repetitive part of the motions to be $\beta$ relaxations. 
At the same time,  disruptions to the repetitive string-like motions by pair-interactions among strings lead to structural relaxations. 
Quantitatively, the probability $P_{ret}$ of hopping reversal reaches 73\% for equilibrium samples and is even higher at lower $T$ for quenched samples. Extended annealing of quenched samples are found to further increase $P_{ret}$. The rate of structural relaxation as determined from particle MSD is close to the rate of non-returning hops but is much smaller than the simple particle hopping rate.

We have proposed that micro-string motions are the elementary dynamical process. Each micro-string requires the presence of a void, which is a quasi particle with a combined free volume comparable to the particle volume distributed among a few interstitial sites. The propagation of the micro-string transports the void from one of its ends to the other.
Recently, we have proposed a distinguishable particle lattice model (DPLM) of glass illustrating these processes in a simplified form \cite{lam2017dplm}. Typical glassy behaviors are reproduced.
\Eq{pret} is in particular strongly supported using equilibrium data.
Moreover, assuming distinguishable particles and disorder quenched in the configuration space, we have derived a local configuration-tree theory of glassy dynamics based on void motions. It quantitatively describes in particular string repetition and interaction behaviors observed here and in DPLM \cite{lam2016}.

\rvs{Properties of string-like motions are remarkably similar in various systems \cite{glotzer1998,glotzer2004,glotzer2003}. 
We expect that the findings in this work based on polymer simulations should apply qualitatively also to other glassy systems exhibiting string-like motions.
In particular, other systems also exhibits hopping dominated by distances comparable to particle separations \cite{wahnstrom1991,sastry1998,kawasaki2013,marcus1999}.   Back-and-forth motions at low temperature are also observed in non-polymeric systems \cite{miyagawa1988,vollmayr2004}.
}
furthermore, the highly repetitive nature of hopping motions is bound to have strong quantitative impacts on the overall dynamics. 
We believe that any detailed theories of glassy dynamics must address or incorporate this finding. 
  
An important early model of dynamic facilitation is the facilitated kinetic  Ising model of Fredrickson and Andersen (FA) \cite{fredrickson1984,fredrickson1985}. In the two-spin facilitation version, a spin can only change state if two neighboring spins are excited as up-spins. An elementary excitation in the form of an up-spin has been interpreted as a fluid region that has a lower than average density \cite{fredrickson1985}.
Our identification of an elementary excitation as the existence of a micro-string which must possess a void is consistent with this picture.
Also, our observation of string interactions in which one string can be enabled or disabled by another string represents facilitation between excitations analogous to those in the FA and related kinetically constrained models (KCM) \cite{ritort2003review} 

We have shown distinct $T$ dependences of the rates $R_1$ and $R_2$ of respectively hops and second non-returning hops. This can be interpreted as a difference in the activation free energies. While the displacement due to a single hop is around $0.9\sigma$,  a non-returning hop generates a larger total displacement around $1.6\sigma$ as explained in Sec. \ref{return}. This is consistent with an activation energy increasing with the displacement magnitude as proposed in Refs. \cite{chandler2011,isobe2016}.

\section*{Acknowledgments}

We thank D.A. Weitz for suggesting the use of a quasi-particle view of void. We also thank many helpful discussions with O.K.C. Tsui and M. Isobe. We are grateful to the support of Hong Kong GRF (Grant 15301014).

\bibliography{polymer_short}

\end{document}